%% file: main.tex
\documentclass[runningheads]{llncs}
\usepackage[T1]{fontenc}
\usepackage{graphicx}

\newcommand{\projname}[1]{\emph{CANTXSec}}

\usepackage[hyphens]{url} %
\usepackage[hidelinks]{hyperref}
\usepackage{multirow}
\usepackage{pifont} %
\newcommand{\cmark}{\yes}%
\newcommand{\xmark}{\no}%
\newcommand{\yes}{\ding{51}}%
\newcommand{\no}{\ding{55}}%
\newcommand{\Yes}{\ding{52}}%
\newcommand{\No}{\ding{56}}%
\newcommand{\soandso}{\textbf{$\sim$}}%
\usepackage{cite} %
\usepackage{subfig} %
\usepackage{tabularx}
\newcolumntype{Y}{>{\centering\arraybackslash}X}
\usepackage[export]{adjustbox}

\usepackage[acronym,shortcuts]{glossaries-extra}    %
\glssetcategoryattribute{acronym}{nohyper}{true}
\setabbreviationstyle[acronym]{long-short}
\input{acronyms}
\usepackage{orcidlink}

\begin{document}
\title{CANTXSec: A Deterministic Intrusion Detection and Prevention System for CAN Bus Monitoring ECU Activations}
\titlerunning{CANTXSec}
\author{
Denis Donadel\inst{1}\orcidlink{0000-0002-7050-9369}  \and
Kavya Balasubramanian\inst{2}\orcidlink{0009-0008-8518-0982}  \and
Alessandro Brighente\inst{1}\orcidlink{0000-0001-6138-2995} \and
Bhaskar Ramasubramanian\inst{3}\orcidlink{0000-0002-4952-5380} \and
Mauro Conti\inst{1}\orcidlink{0000-0002-3612-1934} \and 
Radha Poovendran\inst{2}\orcidlink{0000-0003-0269-8097}
}

\institute{University of Padova, Padua, Italy \and
University of Washington, Seattle, USA \and
Western Washington University, Seattle, USA \\
\email{denis.donadel@phd.unipd.it, kavyab25@uw.edu, alessandro.brighente@unipd.it, ramasub@wwu.edu, mauro.conti@unipd.it, rp3@uw.edu}}
\authorrunning{Donadel et al.}
\maketitle              %
\begin{abstract}
Despite being a legacy protocol with various known security issues, \ac{can} still represents the de-facto standard for communications within vehicles, ships, and industrial control systems.
Many research works have designed \acp{ids} to identify attacks by training machine learning classifiers on bus traffic or its properties. Actions to take after detection are, on the other hand, less investigated, and prevention mechanisms usually include protocol modification (e.g., adding authentication). An effective solution has yet to be implemented on a large scale in the wild. The reasons are related to the effort to handle sporadic false positives, the inevitable delay introduced by authentication, and the closed-source automobile environment that does not easily permit modifying \acp{ecu} software. 

In this paper, we propose \projname{}, the first deterministic Intrusion Detection and Prevention system based on physical \ac{ecu} activations. 
It employs a new classification of attacks based on the attacker's need in terms of access level to the bus, distinguishing between \acp{fia} (i.e., using frame-level access) and \acp{sba} (i.e., employing bit-level access). \projname{} detects and prevents classical attacks in the \ac{can} bus, while detecting advanced attacks that have been less investigated in the literature. 
We prove the effectiveness of our solution on a physical testbed, where we achieve 100\% detection accuracy in both classes of attacks while preventing 100\% of \acp{fia}. Moreover, to encourage developers to employ \projname{}, we discuss implementation details, providing an analysis based on each user's risk assessment.

\keywords{CAN bus \and Intrusion Detection System \and Intrusion Prevention System \and Automotive Security \and Cyber-Physical Systems.}
\end{abstract}
\input{body}

\subsubsection*{Acknowledgments}
This work was supported by the European Commission under the Horizon Europe Programme, as part of the project LAZARUS
(\url{https://lazarus-he.eu/}) (Grant Agreement no. 101070303). The content of this article does not reflect the official opinion of the European Union. Responsibility for the information and views expressed therein lies entirely with the authors. This work was partially supported by project SERICS (PE00000014) under the NRRP MUR program funded by the EU - NGEU. This research was also partially funded by the Research Fund for the Italian Electrical System under the Contract Agreement “Accordo di
Programma 2022–2024 between ENEA and Ministry of the Environment and Energetic Safety – Project 2.1. Moreover, this work has received support from grants N00014-23-1-2386 from US Office of Naval Research and CNS-2153136 from US National Science Foundation.

\bibliographystyle{splncs04}
\bibliography{bibtex.bib}

\newpage
\appendix
\input{appendix}

\end{document}

%% file: acronyms.tex
\newacronym{ai}{AI}{Artificial Intelligence}
\newacronym{asr}{ASR}{Attack Success Rate}
\newacronym{can}{CAN}{Controller Area Network}
\newacronym{cnn}{CNN}{Convolutional Neural Network}
\newacronym{ddos}{DDoS}{Distributed Denial of Service}
\newacronym{dnn}{DNN}{Deep Neural Network}
\newacronym{ecu}{ECU}{Electronic Control Unit}
\newacronym{ev}{EV}{Electric Vehicle}
\newacronym{far}{FAR}{False Acceptance Rate}
\newacronym{gan}{GAN}{Generative Adversarial Network}
\newacronym{gps}{GPS}{Global Positioning System}
\newacronym{gru}{GRU}{Gated Recurrent Unit}
\newacronym{ids}{IDS}{Intrusion Detection System}
\newacronym{lstm}{LSTM}{Long Short-Term Memory}
\newacronym{lstmfcn}{LSTM-FCN}{Long Short-Term Memory Fully Convolutional Network}
\newacronym{ml}{ML}{Machine Learning}
\newacronym{dl}{DL}{Deep Learning}
\newacronym{obd}{OBD}{On-Boad Diagnostic}
\newacronym{rnn}{RNN}{Recurrent Neural Network}
\newacronym{svdd}{SVDD}{Support Vector Domain Description}
\newacronym{svm}{SVM}{Support Vector Machine}
\newacronym{tpm}{TPM}{Trusted Platform Module}
\newacronym{ubm}{UBM}{Universal Background Model}
\newacronym{ubi}{UBI}{Usage Based Insurance}
\newacronym{uds}{UDS}{Unified Diagnostic Services}
\newacronym{mitm}{MitM}{Man-in-the-Middle}
\newacronym{rpm}{RPM}{Revolutions Per Minute}
\newacronym{rf}{RF}{Random Forest}
\newacronym{gmm}{GMM}{Gaussian Mixture Model}
\newacronym{knn}{kNN}{k-Nearest Neighbors}
\newacronym{ips}{IPS}{Intrusion Prevention System}
\newacronym{mcu}{MCU}{microcontroller}
\newacronym{spi}{SPI}{Serial Peripheral Interface}
\newacronym{cps}{CPS}{Cyber-Physical System}
\newacronym{fa}{FA}{False Alarm}
\newacronym{dos}{DoS}{Denial of Service}
\newacronym{fia}{FIA}{Frame Injection Attack}
\newacronym{sba}{SBA}{Single-Bit Attack}
\newacronym{idps}{IDPS}{Intrusion Detection and Prevention System}
\newacronym{ifs}{IFS}{Inter-Frame Spacing}
\newacronym{boa}{BOA}{Bus-Off Attack}
\newacronym{sof}{SOF}{Start-Of-Frame}
\newacronym{crc}{CRC}{Cyclic Redundancy Check}
\newacronym{pcb}{PCB}{Printed Circuit Board}
\newacronym{ics}{ICS}{Industrial Control System}

%% file: body.tex
\section{Introduction}

Nowadays, automobiles are equipped with hundreds of \acp{ecu} controlling vehicle components and functionalities. The de facto standard to allow these devices to communicate is the \ac{can} bus protocol~\cite{can-standard}. 
Officially released in 1986 by Robert Bosch GmbH, it is still employed in almost every vehicle in the world. 
Moreover, it finds applications in other domains such as \acp{ics}~\cite{thompson2018application} and maritime vehicles such as ships~\cite{piketak2009adaptation}.
Despite \ac{can} clear advantages, such as high resilience to faults and the requirements of two wires only~\cite{can-standard}, it suffers from intrinsic security issues~\cite{checkoway2011comprehensive,miller2015remote,serag2022attacks,kulandaivel2021cannon, maggi2017vulnerability}. 
For instance, the absence of authentication allows effortless spoofing~\cite{iehira2018spoofing}, while the fault-tolerant error handling protocol allows \ac{dos} attacks against specific vehicle components~\cite{cho2016error}.  
Tackling these issues with an easily deployable solution is tough due to the sensitivity of networks in safety-critical \ac{cps}, where availability is the main concern. 
Moreover, the diffusion of the \ac{can} bus technology in the automotive industry and beyond makes it difficult for manufacturers to accept radical changes to the standard. %

Over the years, researchers have proposed different solutions to fix security issues, although none are widely implemented at the current time. 
A class of mitigations employs cryptographic primitives to authenticate frames and hide transmitted information through encryption~\cite{halabi2018lightweight,groza2012libra, hartkopp2012macan,nurnberger2016vatican,doan2017can,siddiqui2017secure,lotto2024survey}. 
However, these approaches suffer from several drawbacks. 
While it is difficult to deploy authentication mechanisms supporting retro-compatibility with already employed devices~\cite{doan2017can}, it is even more complicated to imagine imposing a standard that mandates legacy \acp{ecu} to manage cryptographic functions, effectively making all \acp{ecu} already on the market unusable.
Moreover, the close source automotive environment makes it cumbersome to reverse proprietary frame formats and reduces the number of people that may get involved~\cite{pese2019librecan}.
As the \ac{can} bus serves as a safety-critical communication channel, it is imperative to control delays tightly. Cryptographic solutions typically influence performance, even if symmetric encryption can minimize them~\cite{zhang2019improving}, thus presenting limitations.

To deal with the limitations of this environment, researchers started looking at it from different angles, implementing solutions working on top of the untouched bus. 
A classic approach is to employ an additional node to monitor the network and detect attacks~\cite{lokman2019intrusion}. 
It results in \acp{ids} that monitor the format and content of frames to detect anomalies~\cite{taylor2015frequency, seo2018gids, desta2022rec, song2020vehicle} or match signatures of known attacks~\cite{jin2021signature, kruegel2003using}. 
Researchers proposed solutions employing voltage-based security systems that monitor the precise voltage each \ac{ecu} imposes on the bus to identify the transmitter~\cite{schell2020valid, xu2019voltage}. 
These solutions require no additional computational costs for \acp{ecu} but have been proven vulnerable to attacks~\cite{bhatia2021evading, sagong2018exploring}.

The inability to autonomously stop attacks is a considerable limitation of \acp{ids}, which need to rely on external systems or human action. 
Countermeasures to attacks are often developed ad-hoc, presenting mitigations to newly discovered attacks~\cite{serag2021exposing,kulandaivel2021cannon,takada2019counter,de2022canflict}. 
However, applying several defense mechanisms could be expensive for a manufacturer, who may need to deal with incompatibilities. 
Moreover, it does not guarantee that the vehicle will be completely secure from attacks. 
Another approach is represented by \acp{ips}, offering the capabilities to detect and react to attacks, preventing or stopping entire categories of threats. 
Unfortunately, there has been limited research investigating \acp{ips} on the \ac{can} bus~\cite{matsumoto2012method,de2021efficient,serag2023zbcan}. 
Although attacks can be effectively prevented~\cite{palanca2017stealth}, this could be explained by the dangers of blocking actions incorrectly identified as attacks (false positives), which may endanger the safety of the vehicle and its passengers. 
Moreover, in the automotive scenario, false positives also represent a problem in a detection-only paradigm. 
Even a low number of false positives may result in alerts being ignored by drivers or several stops to have the vehicle unnecessarily inspected, deteriorating the reputation and trustworthiness of the manufacturer. 
Therefore, developing a comprehensive solution that is unaffected by detection error is essential in this field.

Most of the attacks researchers proposed are based on \ac{can} protocol-compliant attacks, exploiting the bus structure and the error handling mechanisms~\cite{cho2016error, iehira2018spoofing, serag2021exposing}. 
Less attention has been directed to attacks exploiting actions that do not conform to the \ac{can} specification. 
Connecting a malicious device to a vehicle's bus~\cite{headlights} gives the attacker complete freedom in their actions, even injecting single bits instead of entire frames~\cite{cant}. 
Despite ad-hoc devices, even standard compromised \acp{ecu} may suffer from design flaws that make launching these kinds of attacks possible~\cite{kulandaivel2021cannon}. 
If the offensive side of these attacks has been little investigated, to the best of our knowledge, the detection of such attacks has seldom been discussed. 

There is, hence, the need for a solution that overcomes current limitations. 
It needs to offer retro-compatibility without increasing the computational burden on legacy \acp{ecu}, whil at the same time achieving no false positives.
This makes it difficult to employ \ac{ml} models or other probabilistic approaches that are inherently open to rare but nonetheless present false alarms. 
Employing physical properties of the \ac{cps} allows the development of deterministic security solutions resistant to false positives.
While an attacker can easily craft frames~\cite{tindell_can_2019,iehira2018spoofing}, mimicking physical properties is at least harder than imitating network behaviors, if not impossible at all.

\paragraph{Contributions.}
In this paper, we bridge these gaps by proposing \projname{}, the first deterministic \ac{idps} in the \ac{can} bus based on the co-presence of activity on the transmitting line of an \ac{ecu} and one of its frame on the bus.
Specifically, after building a list of message IDs in the bootstrap phase, they are linked to the \ac{ecu} activated when each ID is transmitted in the bus. Then, we employ this list to detect malicious traffic.
Our approach relies on deterministic comparisons, representing a unique and essential improvement over probabilistic \ac{ids} and \ac{ips} in the literature (e.g., \ac{ml}-based~\cite{minawi2020machine, seo2018gids, de2021efficient}), virtually getting rid of any false positives. Moreover, while an extra effort related to wiring is needed, \projname{} does not require any modification to the standard, resulting in a solution compatible with legacy devices.
It achieves 100\% detection and prevention of common attacks compliant with the \ac{can} standard discussed in the literature while being able to precisely detect advanced attacks based on deviations from the protocol specifications.

Our contributions can be summarized as follows:
\begin{itemize}
    \item %
    We propose a new classification system to separate known and future \ac{can} bus attacks between classical \acp{fia} (i.e., using frame-level access) and advanced \acp{sba} (i.e., employing bit-level access). The two classes represent different access levels by the attacker and can be employed to classify future \acp{ids} in the field.
    
    \item We introduce \projname{}, a deterministic %
    \acrfull{idps} in the \ac{can} bus based on the co-presence of frame IDs on the bus and corresponding \ac{ecu} activations. %
    Our solution only requires cheap additional hardware, without any software modifications, cryptographic protocol, or time-consuming training processes, while preventing \ac{fia} attacks and detecting \ac{sba}, zeroing out false positives.
    
    \item Through the development of a real-world testbed, we demonstrate that \projname{} achieves 100\% accuracy in detecting both categories of attacks. Moreover, we prove that our system can prevent \ac{fia} with a success rate of 100\% without disrupting other communications in the bus. The deterministic detection ensures that no legitimate frames are stopped, thus making \projname{} applicable in commercial devices without risks.
    
    \item We provide an analysis to facilitate the implementation of \projname{} based on each user's risk assessment. In particular, we show that monitoring the activation of 30\% of the overall number of \acp{ecu} in a car ensures attack detection on the majority of safety-critical functions. 
\end{itemize} 

\paragraph{Organization.} The paper starts by providing background insights in Section~\ref{sec:background}. Section~\ref{sec:attacks} describes and categorizes attacks in the \ac{can} bus, while Section~\ref{sec:threat} depicts the threat model we consider in our paper. Then, Section~\ref{sec:ourwork} describes \projname{}, our \ac{idps}, and discusses its implementation. Section~\ref{sec:testbed} illustrates the testbed we employed to obtain the results we discuss in Section~\ref{sec:results}. Relevant related works are presented in Section~\ref{sec:related}, while in Section~\ref{sec:discussion}, we analyze \projname{} with respect to other papers and discuss implementation details. Section~\ref{sec:conclusions} concludes the paper with some final insights.   

\section{Background}\label{sec:background} %
In this section, we discuss some background topics needed to understand the rest of the paper. We overview the \ac{can} bus protocol in Section~\ref{subsec:canbus} and introduce the architecture of \acp{ecu} in Section~\ref{subsec:ecu}.%

\subsection{CAN bus}\label{subsec:canbus}

\ac{can} is a broadcast-based bus employed in \acp{cps} and, in particular, in the automotive environment~\cite{can-standard}. 
The bus works as a logical \emph{AND}: dominant values (zeros) win over recessive values (ones).
After transmitting a bit, each node senses the bus to ensure the transmitted value is actually on the bus. 
If not, a collision is identified. 

\ac{can} frames start with a \ac{sof} bit followed by the ID, which is used to identify message content and as an arbitration mechanism to decide which node is allowed to transmit. 
Lower IDs correspond to higher priority. During the ID transmission, each node sends one bit at a time and senses the bus immediately afterward. 
If a node sending a one senses a zero, it loses the arbitration and thus stops transmitting. 
Through this mechanism, \ac{can} guarantees that, beyond the ID, only one node is allowed to transmit. 

Whenever a collision is identified after the ID, an error is raised. 
Every node keeps two error counters. A Trasmitter Error Counter (TEC) counts errors during transmission, while a Receiver Error Counter (REC) monitors reception errors. Every transmitting error increases TEC by $8$, while receiving error increases REC by $1$. Every correctly sent or received frame decreases the respective counter by 1. Upon encountering an error, the node signals it broadcasting an \textit{error frame}. 

Error frames have different aspects based on the \textit{error state} of the node. Error states are defined by TEC and REC values. Nodes start in the \textit{error-active state}. Once their REC or TEC exceeds $127$, they enter the \textit{error-passive state}. If the TEC exceeds 255, they enter the \textit{bus-off state}, where they stop communicating with the bus. 

When in active error state, error frames are called \textit{active} and are composed of $6$ dominant bits. 
On the other hand, during the passive error state, \textit{passive} error frames are composed of $6$ recessive bits. 
This represents the only occasion when a stream of 6 identical bits could be transmitted in the bus. 
During the transmission of other frames, this is not possible because of \textit{bit stuffing} mechanism. 
It states that whenever a node sends five bits of the same logic level (dominant or recessive), it must send one bit of the opposite level. This extra bit is automatically removed by receivers. This process helps ensure continuous synchronization of the network.

\begin{figure}[tbh]
    \centering
    \includegraphics[width=.8\columnwidth]{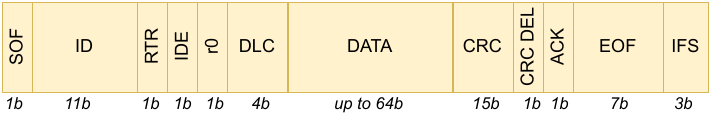}
    \caption{A standard \ac{can} frame showing the bit size of each field.}
    \label{fig:canframe}
\end{figure}

Figure~\ref{fig:canframe} illustrates the different sections of a \ac{can} frame. After SOF and ID, two control bits are sent: the Remote Transmission Request (RTR) indicates remote frame requests, while the Identifier Extension Bit (IDE) signals extended IDs. Then, a dominant reserved bit (r0) is sent. Follows the Data Length Code (DLC), a 4-bit value indicating the length of the data in bytes. Then, the actual content is sent, followed by a \ac{crc} and a recessive bit used as a delimiter (CRC DEL). Then, the transmitter sends a recessive Acknowledge (ACK) bit, during which receivers can confirm the reception of the packet by imposing a dominant value. After the ID, this is the only bit where a node that lost the arbitration is supposed to send data. Finally, seven recessive bits called End-of-frame (EOF) conclude the frame. Before the start of a new frame, three recessive bits are sent and are called \ac{ifs}.

\subsection{Electronic Control Unit}\label{subsec:ecu}
\begin{figure}[tbh]
    \centering
    \includegraphics[width=.7\columnwidth]{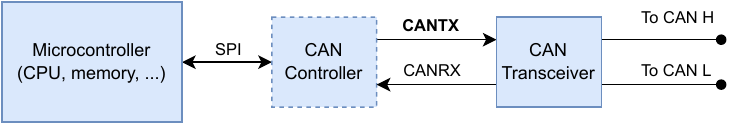}
    \caption{The architecture of a standard \ac{ecu}. 
    }
    \label{fig:ecu_arch}
\end{figure}

An \ac{ecu} is an embedded system employed in automotive electronics to control one or more of the electrical systems in a vehicle. It is usually connected to a variety of sensors and actuators. The process running on an \ac{ecu} varies from simple and well-defined tasks (e.g., a brake system) to more complex and power-consuming operations (e.g., the infotainment \ac{ecu}). 
The majority of \acp{ecu} need to communicate with each other to exchange information about the vehicle's state through the \ac{can} bus.
\acp{ecu} are therefore required to comply with the \ac{can} standard~\cite{can-standard} to allow this communication channel to work properly.
A controller is usually employed as an interface between the \ac{ecu}'s \ac{mcu} and the bus, as shown in Figure~\ref{fig:ecu_arch}. It implements the standard and is in charge of all the \ac{can} related operations, such as buffering and queuing \ac{can} frames that need to be transmitted, packet filtering, arbitration management, and error handling, including the management of error counters~\cite{mcp2515}.   
The controller can be included in the \ac{mcu} \ac{pcb} or added as an external device~\cite{palanca2017stealth}. In this latter case, the \ac{mcu} employs general-purpose communication protocols (e.g., \ac{spi}) to communicate with the controller. This approach allows developers to increase the range of \acp{mcu} they can employ, including boards without \ac{can} bus support.

The logical output of the controller needs to be converted to the differential signaling employed by the \ac{can} protocol. A transceiver converts the controller output (i.e., CANTX) to a signal compliant to the \ac{can} standard~\cite{can-standard}. Moreover, it reads bits in the bus and transmits values to the controller through the CANRX wire~\cite{SN65HVD230}. While it is possible to implement the controller as software~\cite{cant}, the transceiver is essential to convert digital values to signals that can be understood by the bus.    

\section{Attacks on CAN bus}\label{sec:attacks}
\begin{table}[tbh]
\centering
\caption{Attacks available in the \ac{can} bus in the literature and effectiveness of detection (Det.) and prevention (Prev.) of \projname{}. $^1$: detectable for packets with spoofed IDs; $^2$: starting from the first spoofed packet.} %
    \begin{tabular}{l|c|cc} \hline
    \textbf{Attacks}                                  & \textbf{Type}     & \textbf{Det.} & \textbf{Prev.} \\ \hline
   Flooding/DoS~\cite{park2023flooding}              & \ac{fia}   & \yes                 & \yes              \\
    Frame Spoofing~\cite{tindell_can_2019}            & \ac{fia}   & \yes                 & \yes                  \\
    Adaptive Spoofing~\cite{tindell_can_2019}         & \ac{fia}   & \yes                 & \yes                  \\
    Replay Attack~                                    & \ac{fia}   & \yes$^1$             & \yes$^2$              \\
    Original Bus-off Attack (BOA)~\cite{cho2016error, dagan2016parrot} & \ac{fia}   & \yes                 & \yes                  \\
    Stealthy BOA ``single-frame''~\cite{serag2021exposing} & \ac{fia}   & \yes & \yes \\
    Error Passive Spoofing~\cite{elend_cyber_2017}    & \ac{sba} & \yes                 & \no                  \\
    Double Receiving Attack~\cite{tindell_can_2019}   & \ac{sba} & \yes                 & \no                   \\
    Stealthy BOA ``one-packet''~\cite{serag2023zbcan}  & \ac{sba} & \yes                 & \no                   \\
    Freeze Doom Loop Attack~\cite{tindell_can_2019}   & \ac{sba} & \yes                 & \no                   \\
    Shutdown via Clock Gating~\cite{kulandaivel2021cannon}        & \ac{sba} & \yes & \no  \\
    Selective DoS~\cite{palanca2017stealth}           & \ac{sba} & \yes                 & \no          \\ \hline        
    \end{tabular}%

\label{tab:attacks}
\end{table}

Being employed in safety-critical systems, the \ac{can} bus has been the target of numerous attacks, both from researchers and from malicious actors%
~\cite{cho2016error, serag2023zbcan, serag2021exposing, headlights, tindell_can_2019, kulandaivel2021cannon, miller2015remote}. Table~\ref{tab:attacks} summarizes most of the attacks documented in the literature targeting \ac{can}, indicating the type and the ability of \projname{} to detect and prevent them.

In usual threat models in the literature~\cite{bhatia2021evading, checkoway2011comprehensive, foster2015fast, koscher2010experimental, serag2023zbcan}, the attacker gains access to an \ac{ecu}, which is employed to attack the bus. 
However, \acp{ecu} are not all the same. They can be deployed in different subnetworks while being connected to different sensors and actuators. Moreover, different \acp{ecu} are equipped with different capabilities that could allow attackers to compromise different parts of the system. The \ac{ecu}'s firmware and its hardware architecture are also part of the game: experienced attackers may find and exploit bugs in \acp{ecu} through reverse engineering that results in sophisticated access to the \ac{can} bus, while unskilled attackers may rely only on high-level access to the bus provided by the \ac{ecu} frontend.
All of these factors are critical in assessing the types of attacks that can be carried out and, consequently, the effectiveness of identification and prevention methods.
In this paper, we introduce a distinction between two fundamental attack classes in the \ac{can} bus: \textit{\acrfull{fia}} and \textit{\acrfull{sba}}. 

\paragraph{Frame Injection Attacks.}
\acp{fia} include the most common attacks in the literature, where requirements on the compromised \ac{ecu} are relaxed since the attack does not require any particular capability other than the ability to ask the controller to send frames. 
A malicious entity can obtain such access by compromising an \ac{ecu} with malware~\cite{iqbal2019towards}, exploiting software bugs~\cite{miller2015remote}, or installing a malicious device on the bus~\cite{headlights}. %

Many attacks can be adapted from the IT scenario within this threat model. Flooding attacks with low IDs (i.e., high priority) slow down periodic frames and may lead to \ac{dos}~\cite{park2023flooding}.
Through frame spoofing, an attacker can send tampered information on the bus, possibly taking care of avoiding collisions with other frames with the same ID using an adapting spoofing attack~\cite{tindell_can_2019}.
Network captures from the bus can also be replayed later and several times to hide malicious activities or masquerade other attacks performing a replay attack. 
Moreover, the original \ac{boa} does not require any special access to the bus~\cite{dagan2016parrot, cho2016error}. \ac{boa} exploits the error handling mechanisms to increase error counters in a victim \ac{ecu}, pushing it to a bus-off state and effectively stopping the device from sending any frames. The main challenge is related to synchronization because the attacker's message must overlap with the victim's one. Researchers found different ways to synchronize the attacker device with the bus without any special access to the bus, such as exploiting the periodicity of \ac{can} bus messages~\cite{cho2016error}, the knowledge of the legitimate transmitter~\cite{dagan2016parrot}, or the preceding IDs~\cite{serag2021exposing}. 

\paragraph{Single Bit Attacks.}
If these kinds of \ac{ecu} are the most spread, there exist cases where the attacker needs more \emph{fine-grained access} to the bus, monitoring it not frame-by-frame but bit-by-bit. This can be the case when the adversary needs precise synchronization in the bus and the ability to inject single bits during other \ac{ecu} transmissions, possibly disregarding the \ac{can} bus specification~\cite{can-standard}. 
Different ways exist to achieve this: 
1) installing a malicious device developed ad-hoc without a controller, 
2) bypassing or hacking the controller of an \ac{ecu}~\cite{kulandaivel2021cannon}, or 
3) obtaining access to an \ac{ecu} which does not employ a controller and which, therefore, allows fine-grained access to the bus. 
Apart from the installation of a malicious device, it is worth noticing that the effort needed to obtain this level of access is considerable, and it is not always possible. While it is theoretically feasible with some \acp{mcu}~\cite{nucleo} that allow multiplexing the \ac{can} controller output in standard GPIO pins, it is more complex if the output pins are not reconfigurable or the controller is connected via serial connections (e.g., SPI). Other strategies may be possible by exploiting wrong design choices or by interfering with advanced \ac{ecu} functions such as the clock~\cite{kulandaivel2021cannon}.
We call attacks requiring such an access \acp{sba} since, usually, this control on the bus allows attackers to generate errors by injecting a \emph{single bit} in specific instants during the transmission of frames by other \acp{ecu}~\cite{serag2023zbcan, serag2021exposing, kulandaivel2021cannon}.

With such precise access, many other advanced attacks are possible. For instance, Error Passive Spoofing~\cite{elend_cyber_2017} is a two-stage attack that requires 1) forcing an error passive mode victim into generating an error frame and 2) replacing the recessive data and \ac{crc} bits with the spoofed payload. 
A Double Receiving Attack forces the retransmission of a frame by imposing a dominant value in the last EOF bit~\cite{tindell_can_2019}. 
Moreover, direct bus access enables stealthier versions of already discussed \acp{fia}, such as the one-packet \ac{boa}~\cite{serag2023zbcan} or the Selective \ac{dos}~\cite{palanca2017stealth}.
A legacy feature of the \ac{can} bus offers a technique to decrease the bandwidth of the bus and gain time to finish computations for slow \acp{ecu}. It works by imposing a dominant value in the first IFS bit~\cite{tindell_can_2019}. An adversary may exploit it with a Freeze Doom Loop Attack, forcing the dominant bit in a frame and then again to the consequent overflow error frame that will be generated. 
A peculiarity of these kinds of errors is that they do not increase error counters, making the attack stealthier while keeping the bus busy as long as the attack is ongoing. Kulandaivel \emph{et al.}~\cite{kulandaivel2021cannon} proposed an attack to shutdown \acp{ecu} exploiting design errors of certain \acp{ecu} in the wild.    

\vspace{3pt}
Since our approach is tied to the physical aspects of the attack, we decided to separate attacks into \acp{fia} and \acp{sba}. As extensively described in Section~\ref{sec:ourwork}, \projname{} can detect both kinds of attacks, but in two different ways strictly depending on the category. On the other side, using our approach, prevention is possible for the former category only. Moreover, this categorization is important for future work on risk assessment. Even though it may happen that \acp{sba} have more severe effects, they are usually less likely to happen because of the challenging environment needed to carry them out.

From our classification, we intentionally left out \textit{modification attacks}, where an attacker compromises an \ac{ecu} and uses it to replace the content of legitimate frames, i.e., without spoofing another \ac{ecu}'s ID. Being an attack on the data level and not on the network level, it is independent of the communication protocol employed and, therefore, not in the scope of this paper. In addition, it is worth noting that attackers usually exploit vehicles through over-exposed \acp{ecu}, such as the infotainment system, which in normal scenarios is not required to send critical messages on the bus. Therefore, to create effective damage, attackers usually need to forge message IDs to spoof the identity of sensitive \acp{ecu}. To defend the system against these attacks, various techniques exist in the literature, even directly targeting \ac{cps}~\cite{mitchell2014survey} or vehicle~\cite{chiscop2021detecting} environments. Promising techniques employ lightweight feature extraction and contextual information to detect anomalies in transmitted data~\cite{canlp2024, kalutarage2019context}.

\section{Threat Model}\label{sec:threat}

In this paper, as common in the CAN bus security literature~\cite{bhatia2021evading, checkoway2011comprehensive, foster2015fast, koscher2010experimental, serag2023zbcan}, we consider a malicious attacker with the capabilities to compromise \acp{ecu} software remotely (e.g., via unsecured over-the-air firmware update mechanisms~\cite{sota-problems}, via Internet~\cite{miller2015remote}) or physically (e.g., exploiting insecure OBD-II \ac{ecu} software update mechanisms~\cite{onuma2018method}). We do not consider an attacker able to swap a legitimate \ac{ecu} with a malicious one, while we increase the attacker's strength including physical attackers able to connect new malicious devices to the bus~\cite{headlights}. Furthermore, compared to similar recent works~\cite{serag2023zbcan},
the attacker can hack or bypass the \ac{ecu}'s controller to act on the transceiver directly and thus on the bus. 
This is trivial for a physical attacker but also possible for remote attackers~\cite{kulandaivel2021cannon}. 
However, this advanced threat model has some consequences on which attacks can be detected and prevented, as we will discuss in Section~\ref{subsec:coverage}. %

\section{\projname{}}\label{sec:ourwork}

\begin{figure*}[bth]
    \centering
    \includegraphics[width=.9\textwidth]{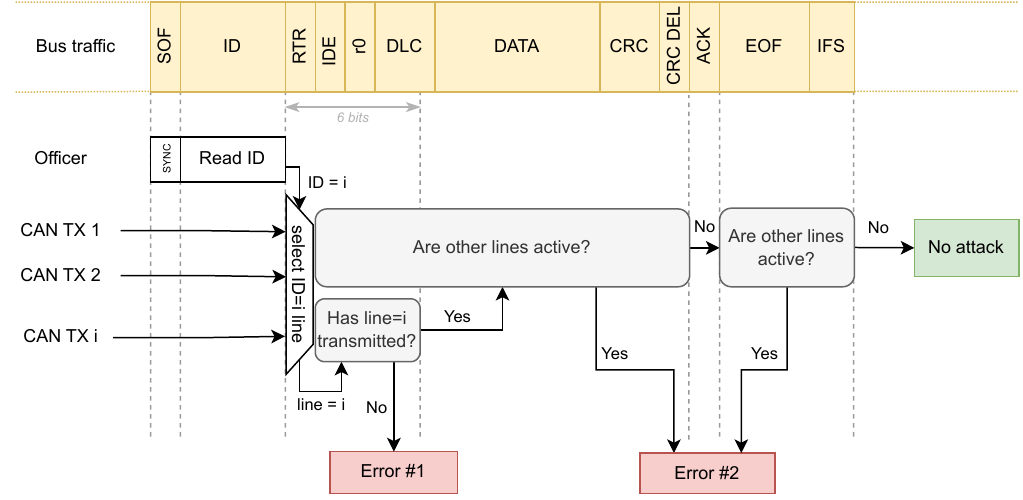}
    \caption{The flow chart explains the different checks \projname{} performs during each bit of every frame sent through the bus. %
    }
    \label{fig:flow}
\end{figure*}

In this paper, we propose \projname{}, the first \ac{idps} on \ac{can} bus, which employs fine-grained measurements from \acp{ecu} activities to detect and prevent network attacks in the bus. \projname{} collects data on the CANTX pin of each \ac{ecu} and correlates these values with a bit-by-bit reading of the bus. When a malicious communication is detected originating from a compromised \ac{ecu} and stopping the frame is safe, \projname{} blocks the transmission of the message by generating errors, eventually forcing the compromised \ac{ecu} to a bus-off state. The architecture of a system employing \projname{} is depicted in Figure~\ref{fig:flow}. A new device, the so-called \emph{officer}, is connected to the bus to detect and react in case of attacks.

\subsection{Officer}
An officer with bit-level access to the bus is the core component of \projname{}. It should be an \ac{mcu} connected to a \ac{can} transceiver. For the officer, a \ac{can} controller is unnecessary since the device should be able to take action on frames almost in real-time without waiting for the end of the packet. To this aim, the officer \ac{mcu} demodulates frames on the fly at the software level. A transceiver is, hence, everything required to enable the \ac{mcu} to interface with the bus if not already provided by the board.

Moreover, the officer is connected via a single dedicated wire to the CANTX lines of all the monitored \acp{ecu} to detect their activities. This wire is only intended to measure digital voltage levels and signal the activation of the CANTX line to the officer. As depicted in Figure~\ref{fig:ecu_arch}, the CANTX line represents the connection between the controller and the transceiver or, if the controller is not present, between the transceiver and the \ac{mcu}. The CANTX line shows a high value when either a recessive bit needs to be sent on the bus or when the transmitter is idle. It instead shows a low value when a dominant bit should be imposed on the bus.

\subsection{Setup}
The first part of the \projname{} setup phase requires physical access to the vehicle and its components. %
In fact, we need to connect a wire from the officer to the CANTX of all the \acp{ecu} to be protected. A perfect time to perform the setup is during automobile manufacturing while assembling the vehicle, to make it easier to reach the various \acp{ecu} and to ensure that no stealthy attacks are ongoing (e.g., malicious \ac{ecu} spoofing IDs during setup). However, a technician can also perform the setup a posteriori, ensuring as much as possible a clean state of the \ac{ecu} software or relying on manufacturer specifications to collect the legitimate IDs. All the \acp{ecu} must be wired to protect against the highest possible number of attacks. 
However, a lot of attacks can be prevented by connected \acp{ecu} even without having all the devices connected to the officer, as we discuss in Section~\ref{subsec:coverage}. Based on the risk assessment conducted on each vehicle, the developer and/or the owner can decide how many and what \acp{ecu} should be protected~\cite{nilsson2008vehicle}.

Each CANTX pin connected to the officer represents a unique \ac{ecu}. For each \ac{ecu}, the officer maintains a list indicating which IDs are allowed to be transmitted from that particular \ac{ecu}. The second step of \projname{} setup is hence the creation of this list. The manufacturer can populate the list by manually inspecting the specifications. Otherwise, it can be achieved by running the automobile in a controlled environment for a sufficient time to ensure that all relevant \acp{ecu} activate at least one time. In the end, the officer should maintain a table where each \ac{ecu} pin corresponds to a list of allowed IDs.
As discussed in Section~\ref{subsec:canbus}, each ID is strictly linked to a particular \ac{ecu}, and there should never be two \acp{ecu} with the same ID in the bus, as per standard~\cite{can-standard}. %

With respect to other \acp{ips}~\cite{de2021efficient}, \projname{} only requires this setup step to start working properly without any time-consuming data collection and training of \ac{ml} models. Updates on the vehicle, such as the replacement or installation of a new \ac{ecu}, require connecting the appropriate cable to the CANTX line and updating the list on the officer. These kinds of actions are straightforward for a technician, and no additional expertise is needed, making \projname{} very easy and fast to upgrade. 

\subsection{Attack Detection}

The intuition on how \projname{} works is as follows. 
Consider a frame with arbitration ID = $i$ being transmitted on the bus. 
For each transmitted bit, we assess whether 1) the correct \ac{ecu} (i.e., the one associated with ID $i$) is transmitting, and 2) all the other \acp{ecu} are quiet. 
Implementing this approach using loops is inconvenient due to the real-time nature of the problem. 
A better solution involves the employment of interrupts.
They are available in modern \acp{mcu} and can be triggered by a voltage change in a pin~\cite{interrupts}.
It is worth recalling that the CANTX lines are set to $1$ while the connected \ac{ecu} is in idle state. 
This makes it more difficult to assess if a certain \ac{ecu} is transmitting a $1$ (recessive value) or waiting. 
However, because of bit stuffing~\cite{can-standard}, a $0$ should be transmitted periodically, even if not included in the original message. 
This makes it possible to detect attacks using edge interrupts with reasonable delays and always within 6-bit time, as discussed in Appendix~\ref{apx:delays}. 
These interrupts fire when a voltage variation is registered in the line, both from a high value to a low value and vice versa.

The \ac{sof} is used by the officer for synchronization, as shown in Figure~\ref{fig:flow}. 
During the arbitration, \projname{} is silent, waiting to know which ID wins the arbitration. 
After arbitration, only the CANTX of the winner \ac{ecu} is authorized to transmit dominant values (except for the Acknowledge bit, as discussed later). 
The officer performs the first check after 6 bits following the end of the arbitration ID. 
In that period, the CANTX line $i$ should have transmitted at least a $0$ (because of bit-stuffing, as detailed in Section~\ref{subsec:canbus}). 
Otherwise, \projname{} raises an Error \#1. 
The second check starts again after the arbitration ID but continues up to the end of the packet, checking if some of the other CANTX lines (i.e., all except for CANTX line $i$) are transmitting a dominant value. 
If so, \projname{} raises an Error \#2. 
This ensures that no \ac{ecu} except the one who wins the arbitration sends bits in the bus.

As depicted in Figure~\ref{fig:flow}, there is one bit that is not considered in this check. In fact, the ACK bit is meant to be set dominant by other \acp{ecu} to acknowledge the reception of the frame. Therefore, the check for Error \#2 is paused during that bit since other \acp{ecu} are allowed to transmit. It is also worth mentioning the existence of a legitimate behavior where the first \ac{ifs} bit is set to zero to signal an overflow. This was used in legacy systems to allow a receiver to delay the next message while executing computation on the previous message~\cite{can-standard}. However, it is no longer employed today and can be exploited to launch a Doom Freeze Attack on the bus~\cite{tindell_can_2019}. Therefore, we do not consider this possibility in our system. However, it can be easily implemented by excluding the check for Error \#2 during that bit.

Compared to other \ac{ids} and \ac{ips} in the literature, \projname{} does not rely on statistical or \ac{ml} models to detect and prevent attacks. Instead, it is based on \emph{physical measurements} and \emph{co-presence} between ID values and corresponding \ac{ecu} \emph{activation} in real-time. This ensures a deterministic solution that leads to perfect attack detection.

\subsection{Attack Prevention}
When the officer detects an attack, it can act in two different ways. One option is just to detect it and alert the driver, who can then decide what action to take. Another option is instead to stop the frame and, therefore, the attack. If detection cannot prevent attacks in real time, it allows alert verification and avoids blocking legitimate packets when false positives are detected. This is one reason why such systems have not found huge implementation in the literature: even if the false positive rate is usually low~\cite{seo2018gids, song2016intrusion, schell2020valid}, it may have an impact on the driving experience. \projname{}, instead, reaches 100\% accuracy without false positives: this allows us to safely implement a prevention mechanism, being sure not to compromise the reliability of the bus.

To stop a frame on the bus, the officer \emph{injects a dominant value until a recessive bit is overwritten}, which happens at most every 6 bits thanks to the bit stuffing mechanisms (see Appendix~\ref{apx:delays}). This generates an error on the transmitter, which stops the frame transmission and starts sending an error frame. The error frame depends on the \ac{ecu} state. In error active mode, it is composed of 6 dominant bits, while in passive mode, it comprises 6 recessive bits. However, even if this is enough to stop the frame~\cite{de2021efficient}, it may not be the smartest strategy to block the attack completely. By default, controllers encountering a transmission error will reschedule the same packet again in the next frame time, which will be stopped by the officer again~\cite{mcp2515}. Such a process increases the busload, and the attacker may exploit this behavior to slow down the bus, performing a \ac{dos}. 

A more efficient strategy aims to force the compromised \ac{ecu} into a bus-off state. While there exist different ways to perform the so-called \emph{bus-off attack}~\cite{cho2016error, serag2021exposing}, the most efficient is the \emph{Instant Bus-Off} proposed by Serag \emph{et al.}~\cite{serag2023zbcan}. It forces an \ac{ecu} into a bus-off state, targeting a single packet in $510 \mu s$ on a $500 kbps$ bus. The idea behind this attack is to target one packet with an error and then target the error frames generated by the first error injection. By iterating this process, error counters on the compromised \ac{ecu} will increase, eventually resulting in a bus-off state. More details are available in the paper~\cite{serag2023zbcan}. 

While \acp{fia} can be easily stopped with this strategy, \acp{sba} cannot. \acp{sba} generate errors while transmitting of another frame by a legitimate \ac{ecu}, violating the \ac{can} protocol~\cite{can-standard}. Since, when the attack is detected, the frame being transmitted is the legitimate one, the same strategy cannot be applied. Otherwise, it will target the legitimate \ac{ecu}. For this reason, the safest solution is only to detect the attack and notify the user, who should decide the most suited action based on the context. Discriminating between attacks to be stopped and attacks to be detected is easy since Error \#1 is related to \acp{fia} that can be prevented, while Error \#2 indicates that a user action is required to avoid jeopardizing the vehicle's safety.

\subsection{Development}

A suitable officer to implement \projname{} should reproduce the logic summarized in Figure~\ref{fig:flow}. Since no \ac{ml} or other computationally expensive calculation is involved, a cheap \ac{mcu} is enough as long as it provides edge interrupts and enough GPIO pins to connect all the monitored \acp{ecu}. Precise analog voltage measurement capability by GPIO pins is not needed since the officer only needs to obtain the digital level of CANTX lines. Moreover, the officer should include a connection to the bus, which allows reading each singular bit in real-time (i.e., without a controller). Because of the controller's absence, the software is responsible for decoding each frame ID and deciding which CANTX lines should be monitored. To identify errors, a computationally lightweight approach includes the employment of interrupts to detect changes in the CANTX lines connected to the officer's GPIO pins. Errors and prevented attacks should be logged and reported to the user so that further analysis can be performed. The code of our prototype and all the devices employed in the testbed are available in our Github repository\footnote{\url{https://github.com/donadelden/CANTXSec/}}.

\section{Testbed}\label{sec:testbed}

\begin{figure}[tbh]\centering
    \subfloat[%
    The officer is connected to the CANTX of the monitored \acp{ecu}, which is the wire connecting the transceiver (T) and the controller (C), when available. \label{fig:testbed_schema}]
    {\raisebox{18pt} %
    {\includegraphics[width=.46\columnwidth]{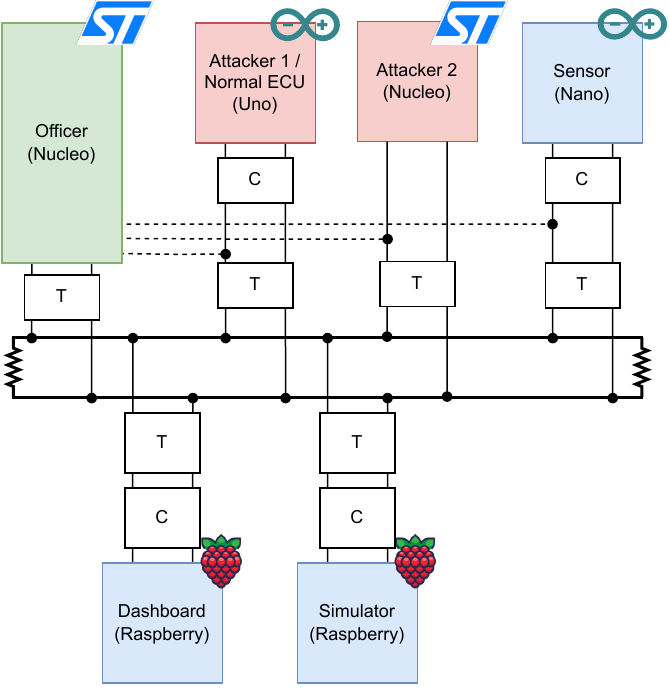}
    }}
    \hfill
    \subfloat[%
    \ding{182}: Dashboard; \ding{183}: Simulator; \ding{184}: Sensor; \ding{185}: Attacker 1 / Normal ECU; \ding{186}: Attacker 2; \ding{187}: Officer.\label{fig:testbed_pic}]{
    \includegraphics[width=0.46\columnwidth]{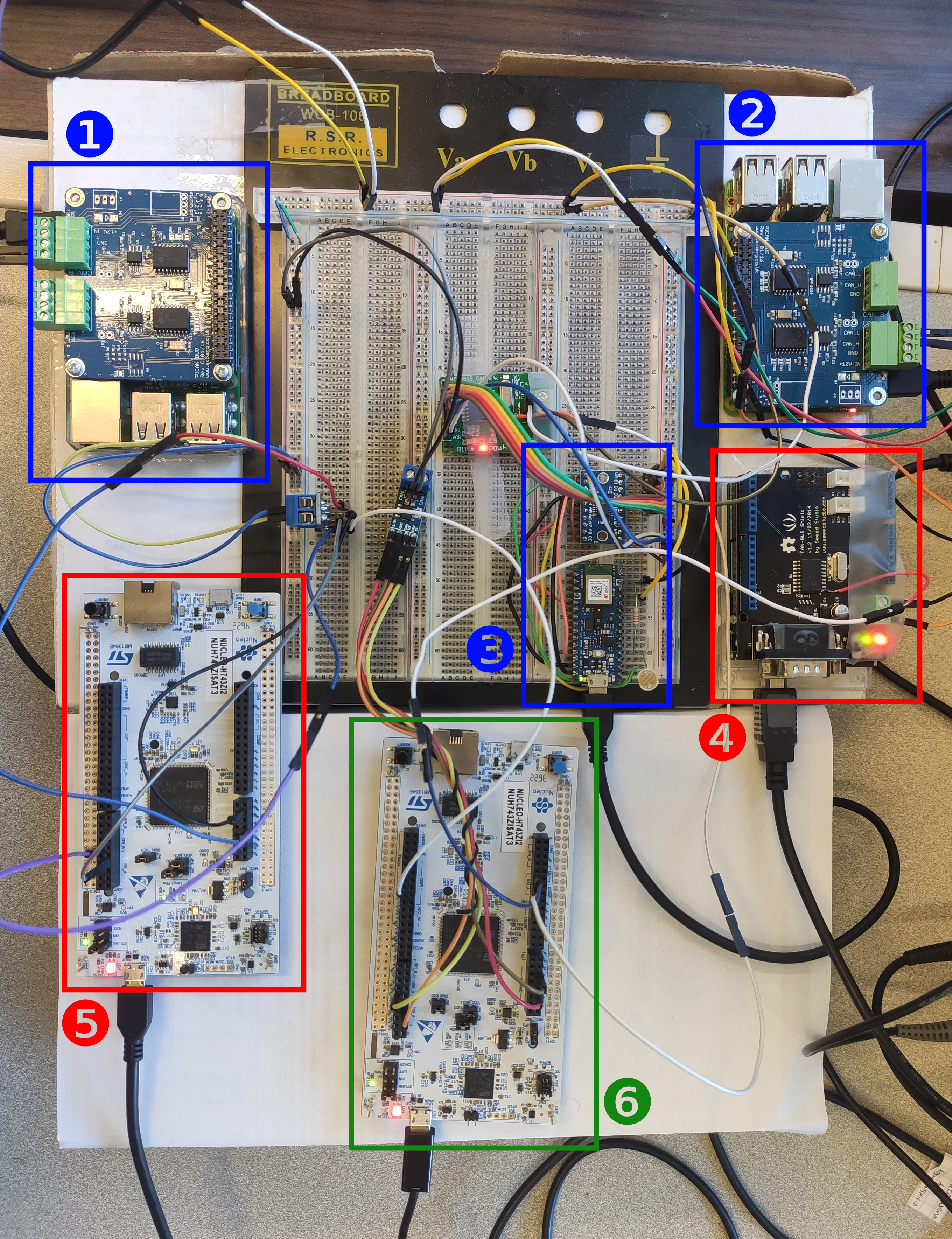}
    
    } 
    \caption{The architecture (Figure~\ref{fig:testbed_schema}) and picture (Figure~\ref{fig:testbed_pic}) of the employed testbed.}

\end{figure}

To test \projname{}, we developed a testbed employing different components to simulate a real \ac{can} bus as closely as possible. We employed different microcontrollers and microprocessors to create a more realistic environment. The testbed schema is depicted in Figure~\ref{fig:testbed_schema}, and the use of each component is described in the following.
\begin{itemize}
    \item \textbf{Officer}: an STM32 Nucleo H743ZI2 board~\cite{nucleo} employed to run \projname{}. It is connected to the bus via a transceiver without any \ac{can} controller and to the \acp{ecu} under control through GPIO pins. 
    
    \item \textbf{Attacker 1/Normal \ac{ecu}}: an Arduino Uno~\cite{arduino} with Seeed Studio shield~\cite{seeedstudio} for the connection to the bus employed as a data transmitter. During the frame spoofing attack, this \ac{ecu} is considered compromised. Otherwise, it behaves as a normal \ac{ecu} broadcasting a random value every 100 milliseconds. The Seeed Studio shield included both a transceiver and a controller. 
    
    \item \textbf{Attacker 2}: an STM32 Nucleo H743ZI2 board~\cite{nucleo} emulating an \ac{ecu} with direct access to the bus (i.e., not mediated by a controller) through a transceiver. %
    It represents the optimal target for an attacker aiming to perform a \ac{sba}.
    
    \item \textbf{Sensor}: developed with an Arduino Nano 33 BLE~\cite{arduino} acts as a sensor broadcasting a physical parameter (the light received by a photoresist) every 100 milliseconds. 
    It represents one of the many sensors a modern vehicle contains to monitor different vehicle and environmental parameters (e.g., the temperature of the engine, external temperature, and tire pressure). 
    
    \item \textbf{Dashboard}: running in a Raspberry Pi 3b~\cite{rpi} with a PICan Duo hat~\cite{picanduo}, it emulates a dashboard, essentially receiving and logging the traffic on the bus. The shield includes both a controller and a transceiver. %
    \item \textbf{Simulator}: running in a Raspberry Pi 3b~\cite{rpi} with a PICan Duo hat~\cite{picanduo}, it created background noise on the bus by continuously sending traffic from a dataset. The dataset has been created by collecting 10 minutes of traffic while using an instrumentation cluster simulator~\cite{simulator}.
\end{itemize}

All the devices are connected to a simulated bus with terminal $120\Omega$ resistors as mandated by the standard~\cite{can-standard}. Figure~\ref{fig:testbed_pic} illustrates the final testbed with all its components.

\section{Results}\label{sec:results}

Even if the attacks presented in Table~\ref{tab:attacks} have different goals, we can summarize them into two categories based on the physical mechanisms needed to perform the attack successfully, as described in Section~\ref{sec:attacks}. Since \projname{} is focused on the bit level, the end goal of the attack is not relevant to the detection. In other words, independently of the final target of the attack, our system is able to identify malicious injections that compose the attack. Therefore, we implemented some representative attacks using different technical methodologies, which we believe are comprehensive for the majority of the attacks in the literature. 

Table~\ref{tab:our_attacks} summarizes the attack's scores and \projname{}'s accuracy in detecting and preventing attacks.
The \ac{asr} is measured depending on the attack's goal, and it is explained in the following sections. The dashboard collects frames successfully sent through the bus while all the relevant \acp{ecu} log the packets transmitted for further analysis and correlation. We ran each experiment for at least 10 minutes, generating several attacks during that timeframe. 
Furthermore, Appendix~\ref{apx:toy} proposes a simple example showing how \projname{} can effectively block an attack and restore the system's normal behavior.

\begin{table}[htb]
    \centering
    \caption{Effectiveness of \projname{} in detecting and preventing the implemented attacks, which are presented with their \acrfull{asr}. Preventing Selective DoS is not possible with this approach.}
    \begin{tabular}{l|cccc} \hline
         \textbf{Attack}& \textbf{\;Type\;} & \textbf{\;\ac{asr}\;} &  \textbf{Detection}&  \textbf{Prevention}\\ \hline
         Frame Spoofing without traffic & \ac{fia} &  100\%& 100\%&  100\%\\
         Frame Spoofing with traffic & \ac{fia}&  100\%& 100\%&  100\%\\
         Selective DoS & \ac{sba} &  100\%& 100\%&  NA\\ \hline
    \end{tabular}
    
    \label{tab:our_attacks}
\end{table}

\subsection{Frame injection}
As a representation of \acp{fia}, we implemented two different frame spoofing attacks, considering that the spoofed ID is also transmitted from another \ac{ecu} or is only sent by the attacker's device. 

\paragraph{Frame spoofing without legitimate traffic.}
In this scenario, the attacker is able to stop a monitored \ac{ecu} from sending data. For instance, they can exploit a \ac{boa}~\cite{cho2016error} or use malware to compromise the \ac{ecu}~\cite{iqbal2019towards}. Moreover, the attack has the control of an \ac{ecu} through which they can send frames on the bus with arbitrary ID. We measure the ASR as the percentage of malicious frames transmitted by the attacker that are correctly received by the dashboard. An attack of this kind has a 100\% ASR since there are no countermeasures applied by default against spoofing attacks. 

Without detection or defense mechanisms, the recipient of the ID would not notice a difference in the packets since the ID is spoofed. If the officer is activated in detection mode, instead, the driver could be notified of every spoofed packet sent on the bus. In the testbed, this happens for 100\% of the attacks, matching with theory since attack detection is deterministic, as explained in Section~\ref{sec:ourwork}. 

\begin{figure}[tbh]
    \centering
    \includegraphics[width=.6\columnwidth]{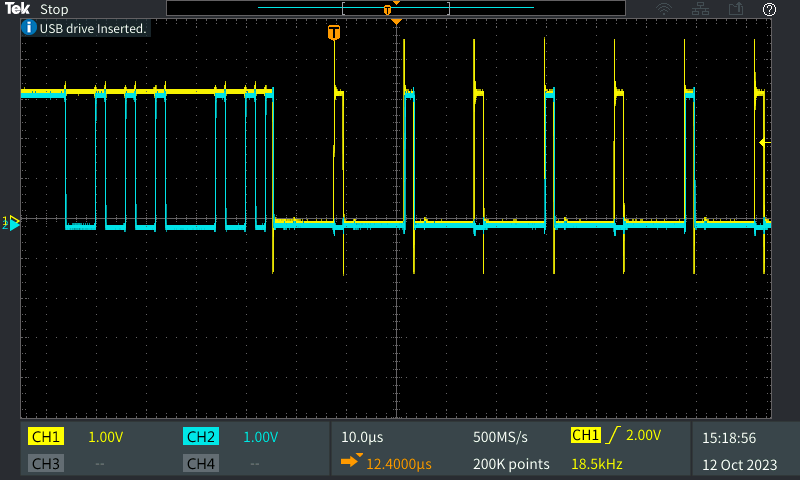}
    \caption{Start of a \ac{boa} as seen in an oscilloscope. The blue signal is the target frame (i.e., the malicious frame). The yellow signal represents the injection made by the officer. The first injection of a dominant value stops the malicious frame. Then, the following injections target error frames automatically generated by the attacker's \ac{ecu}. This increases the attacker's \ac{ecu} error counters, eventually reaching the bus-off state.}
    \label{fig:oscilloscope}
\end{figure}

Furthermore, frame spoofing attacks can be prevented by setting the officer into prevention mode. In this case, when a malicious packet is sensed in the bus, the officer will launch a \ac{boa} against that ID and, therefore, against the compromised \ac{ecu}. Figure~\ref{fig:oscilloscope} shows on an oscilloscope the prevention mechanisms firing against a malicious frame. Even in this case, the success rate is 100\%, and no malicious packets are ever completely sent in the bus. Thus, the receiver does not receive any spoofed data.

\paragraph{Frame spoofing with legitimate traffic.}

Similarly to the previous attack, the malicious actor has compromised an \ac{ecu} through which they can send spoofed frames with any arbitration ID. Compared to the previous scenario, this time, the attacker did not stop the legitimate \ac{ecu} from sending frames on the bus. Therefore, if the attacker sends packets continuously without caring about what is happening on the bus, there may be collisions with the legitimate packets. However, a smart attacker can easily sense the bus and send packets just after legitimate ones~\cite{tindell_can_2019}. With these techniques, they can ensure a 100\% of ASR, which is measured as the percentage of malicious frames correctly received by the dashboard.  

Detecting this attack may seem more challenging since the officer needs to differentiate between malicious and legitimate packets with the same ID. However, since the officer knows the state of the CANTX of the monitored \acp{ecu}, it can tell the source of a frame. With this strategy, during detection mode, 100\% of the malicious packets are detected, while 0\% of the legitimate packets are wrongly classified as malicious. 

In this scenario, prevention shows the same difficulties as detection. When the officer is in prevention mode, it can block all the malicious frames, pushing the attacker's \ac{ecu} into the bus-off state without modifying the behavior of legitimate frames that are received by other \acp{ecu} correctly. In our experiments, \projname{} got a success rate of 100\% in stopping malicious frames.

\subsection{Single bit access attacks}
If detecting and partially preventing frame spoofing has already been investigated by other works, \acp{sba} are intrinsically more challenging to detect. These kinds of attacks bypass the controller with different techniques, for instance, using a modified \ac{ecu}, hacking the controller, or exploiting design errors (e.g., bus clock gating~\cite{kulandaivel2021cannon}). 

As a representative example of these attacks, we implemented the \emph{Selective DoS}, first proposed by Palanca \emph{et al.}~\cite{palanca2017stealth}. Usually, these kinds of attacks exploit the generation of errors in the bus to maliciously stop a packet or increase error counters in \acp{ecu}. Since these bits are injected during the transmission of legitimate packets, trying to stop them by launching \acp{boa} is counterproductive for two reasons. First, this will likely stop the legitimate frame and possibly help the attacker by increasing even more victims' error counters. Second, to control single bits in the bus, the attacker has probably bypassed the controller in some way and, therefore, they no longer need to behave to the standard. For these reasons, we just investigate the detection of these kinds of attacks since prevention is not possible with our strategy. After detection, the victim should stop the automobile and ask for assistance to identify and fix the problem. 

The Selective DoS~\cite{palanca2017stealth} works by injecting a dominant value during the transmission of a recessive value, thus generating an error that stops the packet transmission. The transmitting \ac{ecu} reacts by sending an error frame and rescheduling the transmission of the packet as soon as possible.
Then, the attacker uses the same strategy to stop the retransmitted frame, and the cycle begins again. In some cases, the victim \ac{ecu} may go into the bus-off state, completely stopping packet transmissions.

We implemented the attack on a Nucleo H743ZI2 board~\cite{nucleo} (Attacker 2 in our testbed) by monitoring the bus for frames with the target arbitration ID and then injecting a dominant value during the first transmission of a recessive bit. To check the effectiveness of the attack, we monitor the reception of packets from the dashboard while launching the attack. We measure the ASR as the percentage of packets correctly stopped by the attacker. We assessed that no packets are received when the attack is active, indicating an \ac{asr} of 100\%. 

The detection of Selective DoS goes through the identification of injections of single bits after the transmission of the arbitration ID. Since, by design, the attack will generate a lot of consecutive traffic (i.e., the retransmissions), the officer \ac{mcu} is not fast enough to print an alert for each message. Therefore, we employed a different strategy. We program the attacker to target exactly 200 frames (regardless of whether they are first sends or retransmissions), and we program \projname{} to alert every 200 detection. We avoid notifying each attack on the serial port since this action is time-consuming compared to the attack frequency, and it may result in some attacks passing unnoticed during the notification. By repeating this process several times, we were able to ensure that all the attacks were identified by \projname{}, thus reaching a 100\% detection rate for \ac{sba}. 
In a production environment, there are different strategies to mitigate this issue. First, a faster \ac{mcu} may be employed, possibly implementing the logic on a Field Programmable Gate Array (FPGA) to maximize performances. Second, another \ac{mcu} may be delegated to collect alerts and queue the printing of errors to the driver. Third, since getting the precise number of attacks is usually not essential, the driver may be alerted only for the first attack received.

\section{Related Works}\label{sec:related} %

\paragraph{\ac{ids} on CAN bus.}
Plenty of works in the literature propose to detect attacks in the \ac{can} bus by analyzing frame content employing \ac{ml} and \ac{dl} models~\cite{seo2018gids, song2020vehicle}. Different models have been employed, such as Random Forest~\cite{minawi2020machine}, CNN~\cite{desta2022rec}, GAN~\cite{seo2018gids} and LSTM~\cite{song2020vehicle}. 
GIDS~\cite{seo2018gids} is an \ac{ids} proposed by Seo \emph{et al.}~\cite{seo2018gids} together with a dataset including \ac{dos}, fuzzing (i.e., spoofing of packets with random content and ID), and modification. A GAN is trained with benign data and used to detect attacks with 98\% of accuracy.  
The same dataset is also used by Song \emph{et al.}~\cite{song2020vehicle} to test an \ac{ids} based on different \ac{dl} models achieving higher accuracy. 

Other approaches have been discussed in the literature as well. Five different features have been used by Xiu \emph{et al.}~\cite{jin2021signature} to detect spoofing attacks, including the ID, time interval between packets, and three features related to the frame content (correlation, changing amplitude, and value range). Other works~\cite{xu2019voltage, schell2020valid} try to fingerprint \acp{ecu} based on the voltage level imposed on the bus. However, these mechanisms exploit an average over different measurements during the transmission of the same packet, making them unreliable when dealing with \acp{sba}. Moreover, they have been proven vulnerable to attacks~\cite{bhatia2021evading, sagong2018exploring}. 

Time intervals are another feature employed in the literature to detect attacks in the \ac{can} bus~\cite{song2016intrusion, jin2021signature, taylor2015frequency}. However, these techniques have some limitations and cannot provide protection against the injection of single bits during the transmission of legitimate frames (i.e., \acp{sba}).

\paragraph{\ac{ips} on CAN bus.} Because of the bus nature of the \ac{can} protocol, \acp{ips} are not frequent in the literature. The first to theorize the utilization of the error handling mechanism to prevent the reception of malicious messages have been Matsumoto \emph{et al.}~\cite{matsumoto2012method}. They theorize a modification of \acp{ecu} to include a flag to be set when transmitting so as to enable a security device to send an error if someone else is transmitting in the same period. However, a software and hardware modification of each \ac{ecu} is requested, together with the wiring cost of flags and the extra device. It is suitable for new networks and devices but not for legacy systems. Moreover, the authors did not tackle the implementation challenge.
De Araujo-Filho \emph{et al.}~\cite{de2021efficient} propose an \ac{ips} based on an unsupervised Isolation Forest, which requires some unlabelled training data without attacks in it. Moreover, through the injection of error frames, malicious packets can be stopped. However, only fuzzing and modification attacks are discussed without including any \ac{sba}.
A different solution is called Parrot~\cite{dagan2016parrot}, a spoofing prevention system based on the fact that the legitimate \ac{ecu} being spoofed can easily detect it and launch a bus-off attack against the attacker. However, other attacks, such as \acp{sba}, are not discussed, and some limitations hold, for instance, related to the speed of the \ac{can} transmitter and the impossibility of preventing the first spoofed frame.
Another approach has been proposed with ZBCan~\cite{serag2023zbcan}, a security solution able to authenticate messages exploiting the intra-packet time and stop attacks using a \ac{sba}. However, the implementation requires a new device and a modification of \acp{ecu}' software to include authentication management. 
Moreover, \acp{sba} are discussed only as a countermeasure, and the system cannot detect or prevent them. %

\section{Discussion}\label{sec:discussion}
In this section, we discuss some aspects of our work. We compare \projname{} with other approaches in the literature in Section~\ref{subsec:comparison}. Then, we investigate the \acp{ecu} coverage requirement in Section~\ref{subsec:coverage} and discuss limitations in Section~\ref{sec:limitations}.

\subsection{Comparison with other works}\label{subsec:comparison}
Despite the huge amount of \ac{ids} in the literature, only a few papers describe an effective prevention system that can block attacks in the \ac{can} bus~\cite{de2021efficient, serag2023zbcan}, as described in Section~\ref{sec:related}. Table~\ref{tab:comparison} summarizes the capabilities of our work with respect to other relevant papers in the literature. Detection of \acp{fia} is usually one of the main targets of papers and is thus always satisfied, even if with differences. For instance, papers such as GIDS~\cite{seo2018gids} and Song \textit{et al.}~\cite{song2020vehicle} employ \ac{ml} models to detect attacks. However, they only consider certain specific attacks (i.e., DoS, Fuzzy, Gear Spoofing, RPM Spoofing) that were included in their datasets. Scores vary based on the detected attack, with precisions reaching 99.9\%. Another approach employing voltage levels, VALID~\cite{schell2020valid}, only reaches 99.5\% of accuracy and has been proven vulnerable by certain attacks~\cite{bhatia2021evading, sagong2018exploring}.
Detection of \acp{sba} is, instead, a completely new topic, and, to the best of our knowledge, \projname{} is the first paper in the literature addressing this issue. 

Moreover, our solution is also able to stop attacks, which is instead a topic discussed in just a couple of other works. Compared to our solution, both Parrot~\cite{dagan2016parrot} and ZBCan~\cite{serag2023zbcan} require software modification of each \ac{ecu}, which may be cumbersome when dealing with proprietary software in the automotive scenario. On the other side, the hardware modification required by \projname{} can be applied on every \ac{ecu} by inspecting the \ac{pcb} for the CANTX line, which is easier than dealing with integrity protection and digital signatures applied in all modern \ac{ecu} firmware~\cite{karthik2016uptane, secureCodeSig}. 
Moreover, both approaches can only detect \ac{fia}, while \acp{sba} will pass undetected without \projname{}. While both ZBCan and \projname{} are based on almost deterministic solutions, the approach discussed by De Araujo-Filho \emph{et al.}~\cite{de2021efficient} includes training of \ac{ml} models, resulting in increased setup time and need for computation resources. 
Finally, it is worth mentioning the development of secure \ac{can} transceivers~\cite{secureTransceiver} providing certain security measures such as spoofing and flooding protection. However, they cannot protect against threats that are transparent for the transceiver, such as \acp{sba}, which are still possible with such devices.

\begin{table}[tbh]
\centering
\caption{Our solution compared to other \acp{ids} and \acp{ips} in the literature. \soandso~ indicates a lightweight not-\ac{ml} training process.}
    \begin{tabular}{l|cc|cc|c} \hline
                                                  & \multicolumn{2}{c|}{\textbf{Detection}} & \multicolumn{2}{c|}{\textbf{Prevention}} & \multirow{2}{*}{\textbf{Training}} \\
                                                  & \textbf{\ac{fia}}  & \textbf{\ac{sba}}  & \textbf{\ac{fia}}   & \textbf{\ac{sba}}  &                                    \\ \hline
    GIDS~\cite{seo2018gids}                       & \yes               & \no                & \no                 & \no                & \yes                               \\
    Song \emph{et al.}~\cite{song2020vehicle}            & \yes               & \no                & \no                 & \no                & \yes                               \\
    Xu \emph{et al.}~\cite{xu2019voltage}                & \yes               & \no                & \no                 & \no                & \soandso                           \\
    VALID~\cite{schell2020valid}                  & \yes               & \no                & \no                 & \no                & \soandso                           \\
    Jin \emph{et al.}~\cite{jin2021signature}            & \yes               & \no                & \no                 & \no                & \yes                               \\
    De Araujo-Filho \emph{et al.}~\cite{de2021efficient} & \yes               & \no                & \yes                & \no                & \yes                               \\
    Parrot~\cite{dagan2016parrot}                   & \yes               & \no                & \yes                & \no                & \no                                \\
    ZBCan~\cite{serag2023zbcan}                   & \yes               & \no                & \yes                & \no                & \no                                \\
    \textbf{CANTXSec}                             & \Yes               & \Yes               & \Yes                & \No                & \No        \\ \hline              
    \end{tabular}%

\label{tab:comparison}
\end{table}

\subsection{ECUs coverage}\label{subsec:coverage}

The most significant requirement of \projname{} is the need to wire the \acp{ecu} to the officer, which is somehow a limiting feature in a \ac{cps} like an automobile, while it can be more easily applicable in \acp{ics} or ships. However, it is worth noting that not all the \acp{ecu} in a vehicle have the same risk model, and \projname{} can be easily implemented and deployed to cover only the most critical \acp{ecu}. For instance, it is essential to protect safety-critical \acp{ecu} (e.g., brakes) to ensure reliable operation and to avoid malicious sudden actions that might harm the driver. On the other side, devices such as the infotainment system are less critical and, if compromised, cannot physically harm the passengers. An iteration of such an analysis prioritizes the \acp{ecu} that most require protection.

The number of \acp{ecu} in a vehicle is quite variable. Cheap automobiles are equipped with only a bunch of them, while luxury cars may have up to 150 \acp{ecu}~\cite{ecusInACar}. Accessing information about the number of \acp{ecu} in a vehicle and the related messages is tough because of the closed source of the automotive environment. 
According to openDBC~\cite{openDbc}, a collection of reverse-engineered database of \ac{can} messages for different vehicles, they have a mean of 35 \acp{ecu}, even though the number is quite variable, and databases are not always complete. Usually, each \ac{ecu} sends various messages through different frame IDs, indicating the content of the message and its priority. Since, in the worst-case scenario, attacks target the entire \ac{ecu} (e.g., bus-off attacks will stop the \ac{ecu} from sending any message), in our analysis, we considered only the lowest ID for each \ac{ecu}.

Even though the repository~\cite{openDbc} does not offer a complete collection of vehicle messages and \acp{ecu}, we analyzed the most complete dataset to estimate the number of \acp{ecu} \projname{} should be connected to in different scenarios. Since the ID indicates the priority of the frame, we suppose that lower IDs correspond to safety-critical messages, while high IDs are reserved for infotainment and other non-critical applications. Therefore, to roughly estimate the number of safety-critical \ac{ecu}, we sort them by IDs and threshold them with the first clearly not-critical \ac{ecu}. A utility automobile such as a Hyundai Kia contains 43 \acp{ecu}. Out of them, we identify the $12$ \acp{ecu} with lower IDs to be considered safety-critical, using as a threshold the ID of the parking assistance \ac{ecu}, which can be considered non-critical. More expensive vehicles, such as BMW E9X, contain slightly more \acp{ecu} ($50$). We identify $14$ of them as safety-critical, using as a threshold the IDs adaptive front lighting system, which represents the set of the most important \acp{ecu} to be covered by \projname{}. The effects of limited coverage are discussed deeper in Appendix~\ref{apx:ecu_coverage}.

Based on the simple risk assessment we propose, or others in the literature~\cite{nilsson2008vehicle}, each vehicle owner can decide how many \acp{ecu} should be covered by \projname{}. Just by connecting less than 30\% of \acp{ecu} to the officer, it is possible to protect safety-critical \acp{ecu} from \acp{fia} from both remote and physical attackers. If more than one \ac{can} bus is deployed in a vehicle, another solution could be to cover only the \acp{ecu} connected to the safety-critical buses (e.g., the power train), leaving unprotected less essential networks (e.g., the infotainment systems). While this could be the most widely adopted scenario, high-risk usages may require that all the \acp{ecu} should be covered with \projname{} to ensure full detection of both \acp{fia} and \acp{sba}.

\subsection{Limitations}\label{sec:limitations}

This work introduced for the first time a novel strategy to reduce the long-discussed issue of false positives in the field of attack detection for \acp{cps} and propose the first strategy to eradicate them. However, some limitations should be addressed in future works. 

First, the requirement to cover all the \acp{ecu} with \projname{} to comprehensively detect all the discussed attacks may restrain some producers who could see production prices and vehicle weight rise. However, we offer the reader an analysis to consciously decide how to deploy the system based on the personal risk assessment. Moreover, the most common and straightforward attacks (i.e., \acp{fia}) are detected and prevented with 100\% accuracy just by connecting to \projname{} the most safety-critical \acp{ecu}, relaxing the full wired requirement for the standard user while providing the possibility to increase security for high-risk cases.

Another weakness is that the capability of stopping attacks is limited to \acp{fia}. \acp{sba} represent a huge challenge for prevention since they may be effective after a single injected bit, rendering the time frame to act and stop the attack virtually zero. Mitigation strategies could include firmware and architecture hardening to prevent single-bit access to the bus. Another approach could be to physically prevent \acp{ecu} other than the one winning the arbitration from transmitting on the bus when not allowed, for instance, forcing the CANTX pin to a high value. Moreover, forensics tools may be deployed to secure the system after the attack.

Finally, this research focuses on network-level threats, leaving aside data-level attacks such as modification attacks. \projname{} correlates IDs on the bus with \acp{ecu} activations to spot attacks without looking at the actual data transmitted. 
Due to this architecture, modification attacks are not possible to be detected. 
Further research should be conducted to port the proposed deterministic approach to the data level.

\section{Conclusions}\label{sec:conclusions}

In this paper, we bridged the gap in the detection and prevention of attacks in the \ac{can} bus by proposing \projname{}. Our solution is 1) deterministic, 2) covers advanced attacks never discussed in the literature before, and 3) does not require any software modification of \acp{ecu} and is thus easier to implement. To test and analyze its characteristics, we introduced a novel categorization of attacks in the \ac{can} bus based on the access an attacker needs to carry them out. In fact, we introduced \acp{fia} requiring typical frame-level access and \acp{sba} employing more sophisticated bit-level access. Through the usage of a physical testbed we developed, we demonstrated that \projname{} achieves 100\% accuracy in detecting both \acp{fia} and \acp{sba}. Moreover, we demonstrate how our solution is even able to prevent \acp{fia} with 100\% accuracy.

Future works include further tests through the deployment of \projname{} in a real vehicle and augmenting the number of connected devices. Such an improved testbed will allow for new experiments to test performance and latency under high-load conditions (e.g., heavy traffic on the bus). Increasing the number of \acp{ecu} could also allow testing the practical effects of partial \acp{ecu} cover by \projname{}. The development of such a testbed could also allow considerations about the complexity of deploying \projname{}, such as evaluating the ease of wiring the CANTX lines of legacy and modern \acp{ecu}. Moreover, this work opens a new research direction aiming at the prevention of the newly introduced category of threats we called \acp{sba}, which is a topic not addressed in the literature up to now.

%% file: appendix.tex
\section{Appendix}

\subsection{Attack Detection Delay}\label{apx:delays}

The fastest way for a \ac{mcu} to monitor several lines is through the usage of interrupts. Therefore, we employed changing edge interrupts to check every CANTX line our officer is monitoring. We set the interrupts to fire every time a rising or falling edge is observed on the line. This is perfect for detecting Error \#2, which is generated by an \ac{ecu} passing from idle (recessive value) to a dominant value on the bus. However, it can also be used to identify if the correct \ac{ecu} is transmitting some data. It can be done by monitoring the first bits after the arbitration ID, looking for a bit of change. Theoretically, in the space of binary messages, there exist cases where all the bits in the frames are the same. Practically, this is not possible in a \ac{can} bus because of bit stuffing that forces a different bit after five consecutive equal bits. 

\begin{figure}[tbh]
    \centering
    \includegraphics[width=.6\columnwidth]{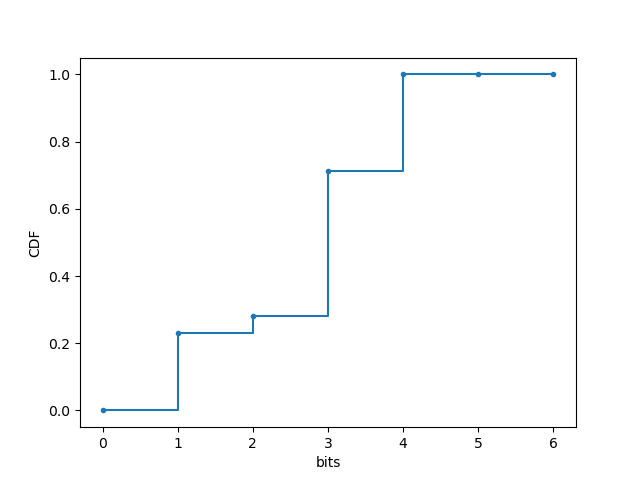}
    \caption{CDF of maximum bit times to wait for an Error \#1 to be raised. }
    \label{fig:wait_times}
\end{figure}

Even if we understand that Error \#1 will eventually trigger, it is interesting to know the mean delay for having such an error raised. The theoretical limit is given by the definition of bit stuffing: $5+1$ bit times. However, we investigated the probability of waiting that time before detecting Error \#1 by computing the time to wait for an edge after the completion of the arbitration ID. We extracted data from a dataset of 10 minutes of \ac{can} traffic to compute the number of bits to wait for an Error \#1 to be raised. Figure~\ref{fig:wait_times} shows the probability of waiting up to a certain amount of bit times to have an alert. As shown, after $4$ bits time, almost all the frames have transmitted a dominant value. However, $6$ bits is the maximum number of bits to wait to have a $0$, as mandated by the bit stuffing mechanism. This guarantees an upper bound on the time to wait for the detection, which is anyway negligible, and ensures detection and prevention of attacks before the end of the frame.

\subsection{A toy example of attack prevention}\label{apx:toy}
To demonstrate the capabilities of \projname{}, we conducted an experiment in a scenario that could happen in a real vehicle. In particular, we imagine an attacker launching a spoofing attack against a sensor in order to tamper with the real value. We targeted a light sensor that is available in our testbed. Similar sensors are everywhere in vehicles, for instance, to measure engine temperatures or tire pressure. As explained in Section~\ref{sec:testbed}, the environment light value is broadcasted every 100ms. The value is almost constant and exhibits small changes through time since the environment brightness does not usually vary with high frequency.

\begin{figure}[tbh]
    \centering
    \includegraphics[width=.6\columnwidth]{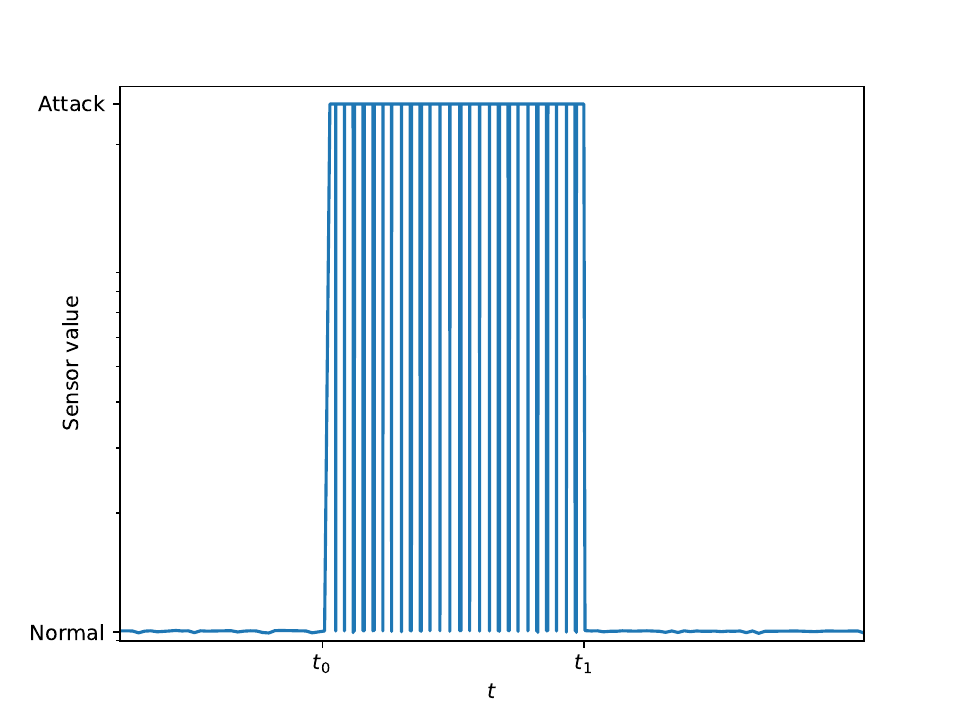}
    \caption{Values of a light sensor over time received by the dashboard. At $t=t_0$, a compromised \ac{ecu} initiates an adaptive spoofing attack, forcing the sensor value to be abnormally high. At $t=t_1$, \projname{} is activated in prevention mode, and the attack is %
    defeated.}
    \label{fig:sensor_attack}
\end{figure}

We started with a bus without any security measures. At the time $t=t_0$, we imagine that a compromised \ac{ecu} monitored by \projname{} starts sending frames at a very high frequency containing a tampered value and spoofing the light sensor ID. The effect is visible in Figure~\ref{fig:sensor_attack}, where it is easy to spot the malicious high value being imposed. In particular, from $t_0$, the flooding of tampered messages with an abnormally high value tries to push the reading out of the normal scenario. The consecutive drops are caused by the legitimate reading that is still periodically sent from the sensor and, therefore, received from the other \acp{ecu}. An advanced adversary could exploit a \ac{boa} against the legitimate sensor to avoid the drops~\cite{cho2016error,iehira2018spoofing}.

To prevent the attack, we activated \projname{} in prevention mode at time $t=t_1$. Blocking malicious frames restores the
normal value, and the attack is stopped. This highlights our system's capabilities to stop spoofing attacks, which are the starting point for all \acp{fia}. Of course, with respect to this example, in a real vehicle, the system will be activated when the vehicle is turned on, so the attack will be stopped from the beginning.

\subsection{Effect of partial ECU covering on CANTXSec capabilities}\label{apx:ecu_coverage}

Covering more or less \acp{ecu} with \projname{} impacts the attacks that can detected and, possibly, prevented. The detection of \acp{fia} is quite resilient to narrow coverage of \acp{ecu} by \projname{}, as shown in Table~\ref{tab:spoofing}. In fact, spoofing is detected and prevented even if the compromised \ac{ecu} is not connected to the security system. This happens because, during spoofing, the legitimate \ac{ecu}'s CANTX will be idle while the officer detects its ID on the bus. It is worth noticing that the malicious \ac{ecu} could be a not monitored \ac{ecu} or a completely new device connected from the attacker to the bus~\cite{headlights}. Therefore, for normal use cases, covering safety-critical \acp{ecu} is enough to secure the system against common \acp{fia}.  

\begin{table}[tbh]
\captionsetup[subfloat]{position=bottom}
\caption{Detectability of a spoofing attack (Table~\ref{tab:spoofing}) and a \ac{sba} (Table~\ref{tab:sba_detect}) for different configurations of monitored or unmonitored \acp{ecu}. $\bar{X}$ means the ECU sending ID=$X$ is monitored by \projname{}.}
\centering
\renewcommand{\arraystretch}{1.5}
\subfloat[Spoofing attack.\label{tab:spoofing}]{
\begin{tabularx}{.35\textwidth}{YY|YYYY}
                                         &                    & \multicolumn{4}{c}{\textbf{Spoofed ID}}                   \\
                                         &                    & \textbf{$A$} & \textbf{$B$} & \textbf{$C$} & \textbf{$D$} \\ \hline
                                         \multirow{4}{*}{\rotatebox[]{90}{\textbf{Origin 
                                         ECU}}} 
                                         & \textbf{$\bar{A}$} & $-$            & \cmark       & \cmark       & \cmark       \\
                                         & \textbf{$\bar{B}$} & \cmark       & $-$            & \cmark       & \cmark       \\
                                         & \textbf{$C$}       & \cmark       & \cmark       & $-$            & \xmark       \\
                                         & \textbf{$D$}       & \cmark       & \cmark       & \xmark       & $-$           
    \end{tabularx}%
}
\quad \quad \quad
\subfloat[\ac{sba}.\label{tab:sba_detect}]{
        \begin{tabularx}{.35\textwidth}{YY|YYYY}
                                           &                    & \multicolumn{4}{c}{\textbf{Victim frame 
                                           ID}}             \\
                                           &                    & \textbf{$A$} & \textbf{$B$} & \textbf{$C$} & \textbf{$D$} \\ \hline
                                           \multirow{4}{*}{\rotatebox[]{90}{\textbf{Attacker
                                           ECU}}} 
                                           & \textbf{$\bar{A}$} & $-$          & \cmark       & \cmark       & \cmark       \\
                                           & \textbf{$\bar{B}$} & \cmark       & $-$          & \cmark       & \cmark       \\
                                           & \textbf{$C$}       & \xmark       & \xmark       & $-$          & \xmark       \\
                                           & \textbf{$D$}       & \xmark       & \xmark       & \xmark       & $-$         
        \end{tabularx}%
}

\end{table}

When dealing with \acp{sba}, the situation is more complicated, as shown in Table~\ref{tab:sba_detect}. In this class of attacks, identifying the transmitter of the malicious bits is essential to detect attacks. All the attacks originated from \acp{ecu} connected to \projname{} are identified since the officer will notice a dominant value during the transmission of a frame not linked to the transmitting \ac{ecu}. However, this does not apply to unmonitored \acp{ecu} since they are outside the control of the officer. This implies that to have complete detection of \acp{sba}, all the \acp{ecu} must be connected to \projname{}.